\documentstyle[12pt]{article}
\newcommand{\be}{\begin{equation}}
\newcommand{\ee}{\end{equation}}
\newcommand{\bea}{\begin{eqnarray}}
\newcommand{\eea}{\end{eqnarray}}
\newcommand{\ba}{\begin{array}}
\newcommand{\ea}{\end{array}}

\def\bbox{{\,\lower0.9pt\vbox{\hrule \hbox{\vrule height 0.2 cm
\hskip 0.2 cm \vrule height 0.2 cm}\hrule}\,}}
\newcommand{\dsl}{\pa \kern-0.5em /}
\renewcommand{\t}{\theta}
\newcommand{\T}{\Theta}

\def\5{\bar }  \def\6{\partial } \def\7{\tilde }
\def\8{\hat }

\newcommand{\beq}{\begin{equation}}
\newcommand{\eeq}{\end{equation}}

\newcommand{\bear}{\begin{eqnarray}}
\newcommand{\eear}{\end{eqnarray}}
\font\mybb=msbm10 at 10pt
\def\bb#1{\hbox{\mybb#1}}

\def\bE {\bb{E}}

\begin{document}


\begin{titlepage}
\vfill
\begin{flushright}
UB-ECM-PF-99/09\\
US-FT/14-99\\
DAMTP-1999-53\\
hep-th/9907022\\
\end{flushright}

\vfill

\begin{center}
\baselineskip=16pt
{\Large\bf Supersymmetric baryonic branes}
\vskip 0.3cm
{\large {\sl }}
\vskip 10.mm
{\bf ~Joaquim Gomis$^{*,1}, $~{}Alfonso V.  Ramallo$^{\dagger,2}$, \\
\vspace{6pt}
Joan Sim\'on$^{*,3}$ and  ~Paul K. Townsend$^{+,4}$ } \\
\vskip 1cm
{\small
$^*$
  Departament ECM, Facultat de F\'{\i}sica\\
  Universitat de Barcelona and Institut de F\'{\i}sica d'Altes
Energies,\\
  Diagonal 647, E-08028 Barcelona, Spain
}\\
\vspace{6pt}
{\small
$\dagger$
 Departamento de F\'\i sica de
Part\'\i culas,  Universidad de Santiago\\
E-15706 Santiago de Compostela, Spain
}\\
\vspace{6pt}

{\small
 $^+$
DAMTP, University of Cambridge, \\
Silver Street, Cambridge CB3 9EW, UK
}
\end{center}
\vfill
\par
\begin{center}
{\bf ABSTRACT}
\end{center}
\begin{quote}
We derive an energy bound for a `baryonic' D5-brane probe in the $adS_5\times
S^5$ background near the horizon of $N$ D3-branes. Configurations saturating
the bound are shown to be 1/4 supersymmetric $S^5$-wrapped D5-branes,
with a total Born-Infeld charge $N$. Previous results are recovered as
a special case. We derive a similar energy bound for a
`baryonic' M5-brane probe in the background of $N$ M5-branes.
Configurations saturating the bound are
again 1/4 supersymmetric and, in the $adS_7\times S^4$ near-horizon limit,
provide a worldvolume realization of the `baryon string' vertex of the
(2,0)-supersymmetric six-dimensional conformal field theory on coincident
M5-branes. For the full M5-background we find a worldvolume realization
of the Hanany-Witten effect in M-theory.

\vfill
 \hrule width 5.cm
\vskip 2.mm
{\small
\noindent $^1$ E-mail: gomis@ecm.ub.es \\
\noindent $^2$ E-mail: alfonso@fpaxp1.usc.es \\
\noindent $^3$ E-mail: jsimon@ecm.ub.es \\
\noindent $^4$ E-mail: p.k.townsend@damtp.cam.ac.uk \\
}
\end{quote}
\end{titlepage}
\setcounter{equation}{0}
\section{Introduction}

Many aspects of the physics of intersecting branes of M-theory or superstring
theory can be understood in terms of the worldvolume field theory of a `probe'
(or `test') brane in the supergravity background of other branes.  One
attractive feature of this approach is that many such supergravity backgrounds
have non-singular horizons near which they are asymptotic to Kaluza-Klein vacua
with an anti-de Sitter (adS) factor \cite{gibtown}; this allows one to make
contact with the physics of supersymmetric conformal field theories (CFT) via
the AdS/CFT correspondence~\cite{malda}. A nice example is provided by a closed
D5-brane `surrounding' $N$ D3-branes  (such that the D3-branes thread the
D5-branes). The Hanany-Witten (HW) effect \cite{HW} implies that each of the
$N$ D3-branes must be connected to the D5-brane by a (`fundamental')
IIB string.
It follows that the D5-brane cannot contract to a point but must instead
contract to some lowest energy configuration involving the D3-branes
and $N$ IIB
strings. In the limit of large $N$ this configuration can be investigated by
replacing the D3-branes by the corresponding supergravity background. As the
D5-brane is allowed to contract (quasi-statically) it enters the near-horizon
region of the background for which the geometry is $adS_5\times S^5$. At this
point the D5-brane is wrapped on the $S^5$ factor and the $N$ strings emanating
from it can be considering as having their endpoints on the $adS_5$ boundary.
This configuration was interpreted in \cite{wit,ooguri} as a `baryon-vertex' of
${\cal N}=4$ $D=4$ super-Yang-Mills (SYM) theory.

Although the above description of the baryon vertex involves IIB strings in
addition to a D5-brane, it is also possible to view it as a single D5-brane
carrying $N$ units of Born-Infeld (BI) charge
\cite{Imamura,callan}. If one assumes that all the BI charge is
concentrated at one point, which may be taken to be a polar point,
then one would expect the minimum energy configuration to be one that
preserves the $SO(5)$ rotational invariance about this point. Such a
configuration is determined by giving the radial position of the
D5-brane in $adS_5$ as a function $r(\theta)$ of the co-latitude angle
$\theta$ on $S^5$. One might also hope that this minimal energy
configuration would preserve some fraction of the D5-brane's
worldvolume supersymmetry. A condition on the function $r(\theta)$ for
preservation of supersymmetry was found by Imamura \cite{Imamura}. An
explicit solution of this condition, which was shown to preserve
1/4 supersymmetry, was then found by Callan et al \cite{callan}.
It was subsequently shown that this solution
saturates an energy bound and hence minimises the
energy for a fixed value of a topological charge \cite{quim}.

One might think that the restriction to $SO(5)$ invariant
configurations is not merely a convenience but
rather a necessity, either for supersymmetry or for minimal
energy. Since the equivalence of these two conditions may not be considered
{\sl a priori} obvious we shall concentrate for the moment
on the minimal energy condition. The minimal energy
configuration with a single pointlike singularity has been interpreted
as the common endpoint of N parallel strings \cite{callan}. For this
interpretation to be valid it must be possible, {\sl given sufficient
energy}, to move the strings apart. In particular,
it must be possible to move their endpoints on the
D5-brane to any point on the 5-sphere such that $r$ becomes a function of all
five angular coordinates of the 5-sphere with N point singularities of unit
charge. It is not guaranteed that a given configuration of this sort will solve
the D5-brane field equations but every such configuration is a possible 
initial configuration satisfying the basic requirement that the
D5-brane carries a total of $N$ units of BI charge. As such it must relax to 
some minimal energy configuration with the same boundary conditions at the N 
singular
points. We shall call these minimal energy configurations `baryonic branes'. 
The issue that we wish to address here is whether the solution of \cite{callan}
is the only baryonic D5-brane or whether it is one of a class of
minimal energy solutions.

We shall attack this problem on several fronts. First, we generalize
the energy bound of \cite{quim} to allow for arbitrary
positions of the $N$ singularities of the radial function on the
5-sphere. The energy is shown to be bounded below by the total charge,
{\sl whose functional form is independent of the positions of the 
singularities}.
The conditions under which this bound is saturated are then shown to be
{\sl precisely those required for preservation of 1/4 supersymmetry}.
This result generalizes the supersymmetry condition of
\cite{Imamura}. Our method, which makes use of
the $\kappa$-symmetry transformations
of the D5-brane \cite{swedes}, differs from that used in
\cite{Imamura}. As for any
super-brane action in any background compatible with $\kappa$-symmetry, the
$\kappa$-symmetry transformations involve a matrix $\Gamma_\kappa$,
characteristic of the background and type of brane. The number of
supersymmetries preserved by the combined background/brane system is the number
of linearly-independent solutions of \cite{bbs}
\be
\label{introa}
\Gamma_\kappa \chi =\chi
\ee
where $\chi$ is an arbitrary linear combination of Killing spinors of the
background\footnote{In all previous applications of this formula it has been
found possible to replace $\chi$ by a constrained but {\sl constant} spinor
$\epsilon$, but this turns out not to be possible for the cases considered
here.}.

Thus, we shall uncover no evidence from either supersymmetry or
Bogomol'nyi-type bounds  that the $SO(5)$ invariant
solution of \cite{callan} is the unique baryonic brane configuration
of either minimal energy or 1/4 supersymmetry. Within the subclass of
D5-brane configurations specified by giving $N$ pointlike
singularities of unit charge at specified points on $S^5$ it is natural to
suppose that the one of minimal energy saturates the energy bound because
there is no apparent obstacle that prevents this. We shall derive an equation
for the function $r$ that is both necessary and sufficient for both saturation
of the bound and 1/4 supersymmetry, so the issue of whether the bound is
saturated boils down to the existence or otherwise of solutions to
this equation subject to specified boundary conditions. For a single point
singularity the equation is solved by the $SO(5)$ invariant configuration of
\cite{callan}. We have found a larger class of solutions, of no particular
symmetry, that depend on all five angular coordinates of $S^5$,
but the function $r$ has branch cuts that make its interpretation
difficult. We have not found any solutions with isolated point
singularities other than the $SO(5)$ invariant one. This may be because they
do not exist, but it could equally well be because they are hard to find. 
This remains an open question.

Another purpose of this paper is to generalize these results for the baryonic
D5-brane to the baryonic M5-brane. To see what this entails, we recall
that the HW effect has an M-theory analogue \cite{HW,deAlwis}. This effect
implies that a closed M5-brane with one
direction parallel to $N$ other M5-branes, and otherwise lying on a 4-sphere at
fixed radius from them, must be connected to each of the enclosed M5-branes by
an M2-brane, with one dimension of the M2-brane in the common direction of all
M5-branes. Let us replace the $N$ M5-branes by a supergravity M5-brane and
consider the remaining one as a probe. Near the horizon of the
supergravity M5-brane the geometry is that of $adS_7\times S^4$ and if
the probe is allowed to contract it will end up as an M5-brane wrapped
on the 4-sphere. This M5-brane is the `baryon string'
vertex of the (2,0) conformal field theory on the $adS_7$ boundary
\cite{ooguri,ali}. It is connected to the $adS_7$ boundary by $N$ M2-branes;
from the $adS_7$ perspective these N M2-branes appear to meet in a string-like
junction at some point in the $adS_7$ interior.

Here we will show that it is posible to view this configuration of
M5-brane and $N$ M2-branes as a single M5-brane carrying $N$ units of
the charge associated with the self-dual closed
worldvolume 3-form. By analogy with the D5-brane case, we will call this
a `baryonic M5-brane'. A baryonic M5-brane wraps the 4-sphere of $adS_7\times
S^4$ with singularities to take account of the non-zero charges. We derive an
energy bound satisfied by such configurations and show that the configurations
that saturate it preserve 1/4 of the bulk supersymmetry. We thereby
find an equation for a function on $S^4$ that is both
necessary and sufficient for either 1/4 supersymmetry and minimal
energy. The general solution of this equation will depend on all four
angular coordinates of $S^4$. We have found explicit solutions of this
type, although they again involve branch cuts and do not have an
obvious baryonic-brane interpretation, except in the subcase of
configurations that are invariant under the
$SO(4)$ subgroup of the $SO(5)$ isometry group of the 4-sphere. These are
specified by a radial function of a single angular coordinate. The requirement
of 1/4 supersymmetry then reduces to a differential equation for
this radial function which turns out to be the same as one recently
found, and solved, for a baryonic D4-brane in a D4-brane background
\cite{callan2, ramallo}.

We shall begin with an analysis of the baryonic D5-brane in
the background of N
D3-branes and then carry out a similar analysis for the baryonic M5-brane in
the background of N M5-branes. The connection with baryon vertices involves
a replacement of these backgrounds by their near-horizon limits but our methods
are general and apply both to the full backgrounds and their near-horizon
limits. The physics of the full background case is, however, rather
different as one cannot expect to
find a {\sl static} D5-brane wrapped around N D3-branes at an
arbitrary distance from them. One instead finds `partially-wrapped'
configurations that provide a worldvolume realization of the HW effect
\cite{callan}. These results have since been extended to all Dp-branes
\cite{ramallo}. Because our analysis of supersymmetry applies
to the full backgrounds  we are able to establish the partial
supersymmetry of these D-brane configurations.

In the M5-brane case neither the near-horizon nor the full M5-brane
background has been previously considered. Our results for the
near-horizon case are as summarized above. In the full background we
again cannot expect to find static $S^4$-wrapped solutions but we can,
and do, find 'partially-wrapped' M5-branes that still preserve 1/4
supersymmetry. These results provide us with a detailed worldvolume
description of the HW effect in M-theory.

\setcounter{equation}{0}
\section{Baryonic D5-brane}

The D3-brane solution of IIB supergravity has all fields vanishing
except the dilaton, which is constant, and the 10-metric and self-dual
5-form field strength $G_{(5)}$. For $N$ coincident D3-branes at the origin,
these take the form

\bea\label{sugrasol}
ds^2_{(10)} & = & U^{-1/2}\,ds^2(\bE^{(1,3)}) + U^{1/2}\left[ dr^2 + r^2
d\Omega_5^2\right]\\
G_{(5)} & = & 4R^4\left[ \omega_{(5)} + \star \omega_5\right]
\eea
where $d\Omega_5^2$ is the $SO(6)$-invariant metric on the unit
5-sphere,
$\omega_5$ is its volume 5-form and $\star \omega_5$ its Hodge dual.
The function $U$ is
\be
U= a + \left({R\over r}\right)^4 \qquad \left(R^4= 4\pi g_s
N(\alpha')^2\right)
\label{harmonic}
\ee
where $g_s$ is the string coupling constant and $\alpha'$ the inverse IIB
string tension. We set $a=1$ for the full D3-brane solution and
$a=0$ for its near-horizon limit.

We now put a probe D5-brane of unit tension into this background such
that it wraps the 5-sphere. Let $\xi^\mu=(t,\t^i)$ be the worldvolume
coordinates, so that $\t^i$ ($i=1,\dots,5$) are coordinates for the
worldspace 5-sphere. The bosonic part of the D5-brane action is
\be
S=-\int_{R\times S^5} d^6\xi
\left[\sqrt{-\det\left(h+ F\right)}
- V\wedge G_{(5)}\right]
\label{action}
\ee
where $h$ is the induced worldvolume metric and $F=dV$ is the
Born-Infeld $U(1)$ field strength. Because of the coupling of $V$ to the
background 5-form field strength $G_{(5)}$, the D5-brane must carry BI electric
charge \cite{wit}. It follows that we cannot take $F$ to vanish
but we can assume
vanishing magnetic charge, in which case $F_{0i}$ are the only non-zero
components of $F$. Following \cite{Imamura,callan,quim}, we shall
now proceed as if we were seeking a worldvolume soliton on the D5-brane,
represented by a (radial) deformation of the D5-brane
in the direction separating it from the backgound D3-branes.
Such a configuration can be represented by the array
\be
\ba{ccccccccccl}
D3: &1&2&3&\_&\_&\_&\_&\_&\_ &\quad \mbox{background}    \nonumber \\
D5: &\_&\_&\_&4&5&6&7&8&\_ &\quad \mbox{probe}     \nonumber \\
F1: &\_&\_&\_&\_&\_&\_&\_&\_&9 &\quad \mbox{soliton.}
\ea
\label{triple}
\ee
The 9-direction is the radial one.

In order to analyze this configuration we choose angular coordinates
$\T^i$ ($i=1,\dots,5$) on the 5-sphere at radius $r$. We then fix
the worldvolume diffeomorphisms by choice of the `static gauge'
\be
X^0 = t\, , \quad \T^i=\t^i \, .
\ee
It now follows from the Hamiltonian formulation of the super D-brane
in a general background \cite{BTtwo} that the Hamiltonian density for
static configurations is given by
\be\label{hsq}
{\cal H}^2 = U^{-{1\over2}}\left[ \tilde E^i\tilde E^j g_{ij}
+ \det g \right]
\ee
where $g$ is the induced {\it worldspace} metric and $\tilde E^i$ is a
`covariantized' electric field density related to $F_{0i}$ by
\be\label{relef}
(\det g) \, F_{0i} = \sqrt{-\det(h+F)}\,  \tilde E^j g_{ij}\, .
\ee
For the background/probe configuration considered here, and in the static
gauge, the Gauss law constraint is
\be
\partial_i\tilde E^i = -4\,R^4 \sqrt{\det \bar g}
\label{gausslaw}
\ee
where $\bar g_{ij}$ is the $SO(6)$-invariant metric on the unit
5-sphere. The source term, due to the non-vanishing 5-form field strength of
the D3-brane background, confirms the statement made above that the D5-brane
carries a non-zero electric charge.

Since we are considering only radial deformations of the D5-brane we
may set
\bea
X^1 = X^2 = X^3 = 0 \, ,
\label{truncation2}
\eea
which leaves the radial function $r$ as the only `active scalar'.
The worldvolume metric for {\sl static} configurations is now
\be
h_{\mu\nu}=\pmatrix{-\,U^{-1/2} & 0\cr 0& g_{ij}}
\ee
where
\be
g_{ij}=U^{1/2}\left(r^2\5g_{ij} + \6_i r\6_j r\right) \, .
\ee
The relation (\ref{relef}) now becomes
\be\label{introlast}
\Delta_5^2 \left(r^2 + \bar g^{ij}\partial_i r\partial_j r\right) F_{0i} =
U\,\left[ \det \left(r^2\5g_{ij} + \partial_i r\partial_j r -
F_{0i}F_{0j}\right)\right]^{1\over2} \left( r^2\bar g_{ij} + \partial_ir
\partial_j r\right)\tilde E^j
\ee
where
\be
\Delta_5=U\,r^4\,\sqrt{\det\5g}\, .
\ee
This implies
\be\label{introlast1}
\tilde E^i\,=\,U^{{1/4}}\,\,
{\sqrt{\det g}\over
\sqrt{1\,-\,U^{{1/2}}\,g^{mn}\,F_{0m}\,F_{0n}}}\,\,
g^{ij}\,F_{0j}\, .
\ee

\subsection{Energy bound}

We now proceed to derive an energy bound for baryonic D5-branes.
We shall do this, following \cite{GGT,quim}, by expressing ${\cal H}^2$ as
a sum of squares. This can be done in more than one way and one gets a
meaningful bound on the energy only if the quantity by which ${\cal
H}$ is bounded is
independent of the quantities neglected in arriving at the bound. This
will be the case if ${\cal H}$ is bounded by the integrand of a
topological charge, and this requirement serves to resolve ambiguities. The
result we obtain by this method will be confirmed in the next subsection by an
analysis of  supersymmetry.

We begin by rewriting the energy density (\ref{hsq}) as
\be
{\cal H}^2 = r^2 \7E^i\7E^j \5g_{ij} + (\7E^i\partial_i r)^2
 +\, (Ur^4)^2 r^2 \det \5g + (Ur^4)^2 \det \5g\, \5g^{ij}
\6_i r\6_j r \, ,
\label{ED}
\ee
This expression is manifestly $SO(6)$ invariant but minimal energy
configurations can be at most $SO(5)$ invariant. We shall later wish
to exploit this fact, so we write the 5-sphere metric as
\be
ds^2 = d\theta^2 + \sin^2\theta\, d\Omega_4^2
\label{smetric}
\ee
where $d\Omega_4^2$ is the $SO(5)$ invariant metric on the 4-sphere,
which we take to have coordinates $\t^{\8{\i}}$.  We may now
rewrite (\ref{ED}) as
\bea
{\cal H}^2 & = & \Delta_5^2 \left(r^2 + (r')^2 +
\5g^{\8{\i}\8{\j}}\6_{\8{\i}} r
\6_{\8{\j}} r\right) + \left(\7E^\t r'\right)^2 +
\left(\7E^{\8{\i}}\6_{\8{\i}}r\right)^2 \nonumber \\
& & + 2\7E^\t r' \7E^{\8{\i}}\6_{\8{\i}}r + \left(\7E^\t r\right)^2 +
r^2 \7E^{\8{\i}} \7E^{\8{\j}}\5g_{\8{\i}\8{\j}}
\label{ED1}
\eea
where the primes indicate derivatives with respect to $\t$.
Note that $\bar g^{\8{\i}\8{\j}}$ are the $\8{\i}\8{\j}$ components of
the inverse $S^5$ metric $\5g^{ij}$.

We can further rewrite (\ref{ED1}) as
\be\label{ED2}
{\cal H}^2 = {\cal Z}_5^2 + \left[\Delta_5\left(r\cos\t \right)'
-\7E^i\6_i \left(r\sin\t\right)\right]^2 +
|\Delta_5 \5g^{\8{\i}\8{\j}}\6_{\8{\j}}r -r\,\7E^{\8{\i}} |^2
\ee
where $||^2$ indicates contraction with $g_{\8{\i}\8{\j}}$, and
\be\label{topo}
{\cal Z}_5= \Delta_5 \left(r\sin\t\right)' + \7E^i\6_i \left(r\cos\t
\right) \, .
\ee
Using the Gauss' law (\ref{gausslaw}), one can show that
${\cal Z}_5=\6_i {\cal Z}_5^i$ where $\vec{{\cal Z}_5}$ has components
\bea
{\cal Z}_5^\t & = & \7E^\t\,r\cos\t +\sqrt{\det\5g}\,\sin\t
\left(a\,{r^5\over 5} + r\,R^4\right) \nonumber \\
{\cal Z}_5^{\8{\i}} & = & \,\7E^{\8{\i}}\,r\cos\t \, .
\label{div1}
\eea
{}From (\ref{ED2}), and the fact that ${\cal Z}_5$ is a
divergence, we deduce the
bound
\be
{\cal H} \geq  |{\cal Z}_5|\, ,
\label{Hbound}
\ee
with equality when
\bea
\7E^{\8{\i}} & = & \Delta_5 {\5g^{\8{\i}\8{\j}}\6_{\8{\j}}r \over r}
\label{BPSe1} \\
\7E^\t & = & {\Delta_5 \over \left(r\sin\t\right)'}
\left(\left(r\cos\t\right)' - {\5g^{\8{\i}\8{\j}}\6_{\8{\i}}r\6_{\8{\j}}
\left(r\sin\t\right)\over r}\right) \, .
\label{BPSe2}
\eea
When combined with the Gauss law (\ref{gausslaw}), these conditions
yield the equation
\be\label{pde}
\partial_{\hat{\i}}\left(\Delta_5\bar g^{\hat i\hat j}
{\partial_{\hat{\j}}r\over r}\right) + \partial_\theta \left[{\Delta_5 \over
(r\sin\theta)'}\left((r\cos\theta)' - \bar g^{\hat i\hat j}\partial_{\hat{\i}}
r\partial_{\hat{\j}}r{\sin\theta\over r}\right)\right] = -4R^4\sqrt{det\,\bar g}
\ee
We shall analyse this equation in more detail later, after rederiving
it as a condition for preservation of 1/4 supersymmetry.

The first-order equations (\ref{BPSe1}) and ({\ref{BPSe2}) have been obtained
for an arbitrary value of the coefficient $a$. As we are mostly interested in
describing the baryonic vertex in the near-horizon region of the background,
we shall take $a=0$ in the remainder of this subsection
(for an analysis of the $a=1$ case see refs. \cite{callan, ramallo}).
{}From the bound (\ref{Hbound}) we deduce the bounds
\be\label{inequality}
H \ge \int d^5 \sigma |{\cal Z}_5\big| \ge
\bigg|\int d^5 \sigma {\cal Z}_5\,\bigg|
\ee
on the total energy $H$. The first inequality is saturated under the
conditions just obtained. The second inequality will be saturated too if ${\cal
Z}_5$ does not change sign within the integration region.
If the D5-brane is to describe a baryonic brane we must take this
region to be the 5-sphere with some number of singular points removed.
For the moment we shall simply assume that the second inequality
{\sl is} saturated when the first one is, in which case the total energy equals
the 4-form dual of $\vec{{\cal Z}_5}$ integrated over the 4-boundary
of the D5-brane. In other words
\be\label{kbound}
H = \lim_{\delta\rightarrow 0} \sum_k \int_{B_k} d\vec{S}\cdot \vec{{\cal Z}}
\ee
where $B_k$ is a 4-ball of radius $\delta$ with the k'th singular point as its
centre. The k'th term in the sum can be viewed as the energy of the IIB
string(s) attached to the k'th singular point. Since we will not be
able to provide an explicit solution of (\ref{pde}) with these
boundary conditions unless $k=1$ we should stress that the right hand side
of (\ref{kbound}) is still a lower {\sl bound} on $H$ {\sl even
if the bound cannot be saturated}.

To make contact with previous work we now specialize to $SO(5)$
invariant configurations. In this case $\tilde E^{\8{\i}}=0$ and
\be
\tilde E^\t = \sqrt{\det g^{(4)}}\, E(\t)
\ee
where $E$ is a function only of $\t$. Similarly, $r$ is now a function
only of $\t$. It will also be convenient to set
\be
\Delta_5 = \sqrt{\det g^{(4)}}\, \Delta
\ee
since $\Delta=\,R^4\,\sin^4\theta$ is also a function only of $\t$.
For such $SO(5)$-invariant configurations the first BPS condition
(\ref{BPSe1}) is trivially satisfied while (\ref{BPSe2}) reduces to
\be
{r' \over r} = {\Delta\sin\t + E\,\cos\t \over
\Delta\cos\t - E\,\sin\t}\, .
\label{oldbound}
\ee
This equation coincides with the BPS equation found previously for
$SO(5)$-invariant configurations \cite{Imamura,callan,quim}. For these
configurations the Gauss law (\ref{gausslaw}) becomes the ordinary first order
differential equation
\be\label{lawgauss}
E' = -4\,R^4\,\sin^4\t \, ,
\label{q1}
\ee
the solution of which is given by \cite{callan}
\be\label{expressE}
E = {1 \over 2}R^4\left[\,\,3\left(\nu\pi-\theta\right)+3 \sin\theta
\cos\theta+
2\sin^3\theta \cos \theta\,\,\right]
\ee
where $\nu$ is an integration constant. As pointed out in \cite{callan},
the symmetry of the problem allows us to restrict $\nu$ to lie in the interval
$[0,1]$.

We now return to consider whether the second inequality of (\ref{inequality})
is saturated when the first one is. We shall restrict our discussion of this
point to the $SO(5)$-invariant case, for which we can write
\be
{\cal Z}_5 = \sqrt{\det g^{(4)}}\, {\cal Z}(\t)\, .
\ee
where ${\cal Z}$ is a function only of $\t$. Using (\ref{oldbound}) we see that
\be\label{expressZ}
{\cal Z}= r\, {\left(\Delta\cos\t -E\sin\t\right)^2 +
\left(\Delta\sin\t + E\cos\t\right)^2\over
\left(\Delta\cos\t - E\sin\t\right)}\, ,
\ee
the sign of which is determined by the sign of the denominator. It follows that
${\cal Z}$ will not change sign as long as it has no singularities within
the integration region $\t\in [0,\pi]$ (except, possibly, at the endpoints
$\t=0,\pi$). Now
\be\label{denom}
\Delta\cos\t - E\sin\t = \,\,
\frac 32 \,R^4\sin\theta
\, \eta(\theta)
\ee
where
\be
\eta(\theta)\equiv \theta-\nu\pi -\sin\theta \cos\theta\, .
\ee

We see from (\ref{denom}) that
the denominator of the expression (\ref{expressZ}) for ${\cal Z}$ vanishes at
the endpoints $\t=0,\pi$ but is otherwise positive provided that
$\eta(\t)$ is positive. For $\nu \in [0,1]$ this condition is satisfied
only for
$\nu=0$. In this case (\ref{expressE}) becomes
\be\label{expressEE}
E = {1 \over 2}R^4\left[3 \left(\sin\theta\cos\theta -\t\right) +
2\sin^3\theta \cos \theta\,\,\right]\, .
\ee
Using this result for $E(\t)$ in (\ref{oldbound}) we find a first order
equation for $r(\t)$ for which the solution is \cite{callan}
\be
r= r_0\left({6\over 5}\right)^{1\over3} ({\rm cosec}\,\t)(\t
-\sin\t\cos\t)^{1\over3}\, ,
\ee
where $r_0$ is the value of $r$ at $\t=0$. It was shown in
\cite{callan} that this configuration corresponds to N fundamental strings
attached to the D5-brane at the point $\t=\pi$, where $r(\t)$ diverges. One may
verify that this is a solution of the second order equation
\be
\left[\,\sin^4\t\,{r'\cos\t-r\sin\t \over
r'\sin\t + r\cos\t}\right]' = -4\,\,\sin^4\t \, ,
\ee
found by combining (\ref{lawgauss}) with (\ref{oldbound}).

The solutions of the first-order equation (\ref{oldbound})  for $\nu\not=0$
have been obtained in  \cite{callan}. In these solutions the range of values of
$\theta$ for which the solution makes sense does not coincide with the interval
$[0,\pi]$ because the D5-brane does not wrap completely the 5-sphere. In fact,
when $\nu\ne0$ the D5-brane reaches the point $r=0$, and can therefore be
interpreted as intersecting the D3-brane.  Because the D5-brane only
partially wraps the 5-sphere it captures only part of the five-form flux. This
is consistent with the fact that the tensions of the D5-brane spikes of these
partially-wrapped solutions correspond to a number of strings less than $N$.
In their study of baryonic multiquark states with $k<N$ quarks in ${\cal N}$=4
D=4 super Yang-Mills theory, the authors of \cite{BI}
consider configurations in
which $k$ strings connect the D5-brane to $r=\infty$ while the remaining $N-k$
strings connect it to $r=0$. It is tempting to think that these configurations
could be related to  the $\nu\not=0$ solutions of the BPS equation.

\subsection{Supersymmetry}

We will now show that configurations saturating the bound
(\ref{Hbound}) preserve
$1/4$ of the bulk supersymmetry. Our starting point is the equation
(\ref{introa}). The matrix $\Gamma_\kappa$ is given in \cite{swedes} as a
(finite) expansion in powers of the BI field strength. Because we assume a
purely electric field strength, this expansion terminates at linear order
and the resulting matrix is
\be
\Gamma_\kappa={1\over 6!}{1\over \sqrt{-\det(h +F)}}
\epsilon^{\mu_1 ...\mu_6}~\left[\Gamma_{\mu_1 ...\mu_6}\sigma_1 +
15 ~F_{\mu_1\mu_2}\Gamma_{\mu_3 ...\mu_6}(i\sigma_2)\right]\, ,
\label{gamma}
\ee
where $F_{0i}$ are the only non-vanishing components of the Born-Infeld
2-form. The matrices $\Gamma_{\mu\nu\cdots}$ are antisymmetrized products
of the
(reducible) induced worldvolume gamma matrices
\be
\Gamma_\mu = \6_\mu X^m E_m{}^{\underline m} \Gamma_{\underline m}
\ee
where $\Gamma_{\underline m}$ are the constant D=10 Dirac matrices
and $E_m{}^{\underline m}$ is the spacetime vielbein. In the static gauge and
for static configurations with $X^1=X^2=X^3=0$ we have
\bea
\Gamma_0 &=&U^{-\,1/4}\Gamma_{\underline 0} \nonumber \\
\Gamma_i &=& U^{1/4}\,r\gamma_i + U^{1/4}\,\6_i
r\Gamma_{\underline{r}}\, .
\eea
Here we have introduced the matrix
\be
\gamma_i = e_i{}^{\underline i}\Gamma_{\underline i}
\ee
where $e_i{}^{\underline i}$ is the $S^5$ f\"unfbein. Thus
\be
\{\gamma_i,\gamma_j\}= 2\bar g_{ij}\, .
\ee
The Killing spinors of the D3-brane background take the form
\be
\chi = U^{-{1\over 8}}\epsilon
\ee
where $\epsilon$ is a covariantly constant spinor on $\bE^{(1,3)}\times \bE^6$
subject to a 1/2 supersymmetry breaking condition imposed by the background,
which we discuss below. The supersymmetry preservation condition (\ref{introa})
is thus reduced to
\be\label{susycondition}
\Gamma_\kappa \epsilon =\epsilon\, .
\ee
Although $\epsilon$ is constant in cartesian coordinates on $\bE^{(1,3)}\times
\bE^6$, it is not constant in the polar coordinates for $\bE^6$ that we are
using here. In polar coordinates a covariantly constant spinor is
independent of the radial variable and hence covariantly constant on $S^5$.
Covariantly constant spinors on $S^n$ were constructed explicitly
in \cite{pope}
for a parameterisation in which the (diagonal) metric is found by
iteration of $ds^2_n = d\t_n^2 + \sin^2\t_nds^2_{n-1}$. The result can
expressed in terms of the $n$ angles $\t^i=(\t,\t^{\8{\i}})$ and the
antisymmetrixed products of pairs of the constant D=10  Dirac matrices matrices
$\Gamma_{\underline i} =
(\Gamma_{\underline \t},\Gamma_{\underline {\8{\i}}})$. For $n=5$ we have,
with the understanding that $\Gamma_{\underline {\hat 5}}\equiv
\Gamma_{\underline\t}$,
\beq
\epsilon\,= e^{{\t\over 2}\,\Gamma_{\underline{r\t}}}\,\,
\prod_{\8{\j}=1}^4\,\,
e^{-{\t^{\8{\j}}\over 2}\,\Gamma_{\underline{\8{\j}{\8{\j}+1}}}}\,\,
\epsilon_0
\label{killing}
\eeq
where $\epsilon_0$ is a constant spinor subject to
\be
\Gamma_{\underline{0123}}(\,i\,\sigma_2)\epsilon_0 =\epsilon_0 \, ,
\ee
which is the 1/2 supersymmetry breaking condition associated with the D3-brane
background. There are additional Killing spinors in the near-horizon limit but
the corresponding background supersymmetries are not preserved by the baryonic
D5-brane probe and can therefore be ignored.

Using the expression (\ref{gamma}) for $\Gamma_\kappa$ we find, after some
simplification, that the supersymmetry preservation condition
(\ref{susycondition}) reduces to
\bea
& U\,\sqrt{\det\left[r^2\bar g_{ij}  + \partial_ir\partial_j r
- F_{0i}F_{0j}\right]}\, \epsilon = \nonumber \\
& \big[ Ur^5  \sqrt{\det\,\5g}\, \Gamma_{\underline
0}\gamma_*\sigma_1 -
Ur^3\sqrt{\det\,\5g}\, F_{0j}\6_i r
\gamma^{ij}\gamma_*\Gamma_{\underline{r}}(i\sigma_2)  \nonumber\\ &
\qquad
+\, Ur^4\sqrt{\det \bar g}\, \gamma^i \gamma_*
\left(F_{0i} (i\sigma_2) + \6_i
r \Gamma_{{\underline 0}{\underline r}}\sigma_1\right)\big]\, \epsilon
\label{kapmatrix}
\eea
where $\gamma^i= \5g^{ij}\gamma_j$ and
\be
\gamma_*=\Gamma_{\underline{45678}}\, .
\ee

We shall seek solutions of this equation for $\epsilon_0$ satisfying the
constraints
\be
\Gamma_{\underline 0}\gamma_*\sigma_1\epsilon_0=\epsilon_0\, ,
\label{susyprobe}
\ee
which is expected from the {\sl local} preservation of 1/2 supersymmetry by the
D5-brane, and
\be\label{bion}
\Gamma_{\underline{0}\underline{r}}\, \sigma_3\epsilon_0= \epsilon_0 \,,
\ee
which is the condition associated with a IIB string in the radial direction.
These relations imply
\bea
\Gamma_{\underline{0r}}\,\sigma_3\,\epsilon\,&=&
\left[\,{\rm cos}\,\t\,-\,
{\rm sin}\,\t\,\Gamma_{\underline{r\t}}\,\right]\,\epsilon \nonumber\\
\Gamma_{\underline{0}}\, \gamma_*\,\sigma_1\,\epsilon\,&=&
\left[\,{\rm cos}\,\t\,-\,
{\rm sin}\,\t\,\Gamma_{\underline{r\t}}\,\right]\,\epsilon
\label{cond1}
\eea
Other useful conditions that also follow are:
\bea
\Gamma_{\underline{i}}\,\,\gamma_*\,(i\sigma_2)\,\epsilon &=&
\,-\,\Gamma_{\underline{ri}}\,\epsilon \nonumber\\
\Gamma_{\underline{i}}\,\,\gamma_*\,
\Gamma_{\underline{0r}}\,\sigma_1\,\epsilon &=&
\Gamma_{\underline{ri}}\,
e^{-\t\Gamma_{\underline{r\t}}}\,\,\epsilon\nonumber\\
\gamma_*\,\,\Gamma_{\underline{r}}\,\,(i\sigma_2)\,\epsilon &=&
\,-\,\epsilon
\label{cond2}
\eea

Using the relations (\ref{cond1}), (\ref{cond2}), one can rewrite the
right hand side of (\ref{kapmatrix}) as
\bea
& \Delta_5\,\bigg[ \left(r\sin\t\right)'
 + \Gamma_{\underline{r\t}}\left((r\cos\t)'-F_{0\t}\right)
+ \Gamma_{\underline{r}}\gamma^{\8{\i}}\left(\6_{\8{\i}}r \cos\t -
F_{0\8{\i}}\right) \nonumber\\
& \, + \gamma^{\8{\i}\8{\j}}\,{1\over r}
\left(\6_{\8{\i}}r F_{0\8{\j}} -\6_{\8{\j}}r F_{0\8{\i}}\right)
+ \gamma^{\8{\i}}\Gamma_{\underline{\t}}\,{1\over r}\left(
\6_{\8{\i}}rF_{0\t}-r' F_{0\8{\i}} + r\6_{\8{\i}}r \sin\t\right)\bigg]\, .
\label{part}
\eea
Requiring the coefficients of $\Gamma_{\underline{r\t}}$ and
$\Gamma_{\underline{r}}\gamma^{\8{\i}}$ in (\ref{part}) to vanish, we obtain
\beq
F_{0i}\,=\partial_{i}\left(r\,\cos\t\right)
\label{bpsbound}
\eeq
which can be interpreted as the BPS equation for a worldvolume BIon. It is
straightforward to check that when (\ref{bpsbound}) is satisfied the
coefficients of $\gamma^{\8{\i}\8{\j}}$ and $\gamma^{\8{\i}}
\Gamma_{\underline{\t}}$ vanish identically, so (\ref{kapmatrix}) is
satisfied as a consequence of (\ref{bpsbound}) provided that
\be
U\,\sqrt{\det\left[r^2\bar g_{ij}  + \partial_ir\partial_j r -
F_{0i}F_{0j}\right]} = \Delta_5  (r\sin\t)' \, .
\label{fin}
\ee
It can be verified that this is identically satisfied when $F_{0i}$ is
given by (\ref{bpsbound}). We have thus shown that (\ref{bpsbound}) is
sufficient for preservation of 1/4 supersymmetry. We believe that it is
also necessary; certainly, it is not difficult to check that there is no
configuration preserving more than 1/4 supersymmetry.

We have now found the condition on $r$ and $F_{0i}$ required for
preservation of supersymmetry. Inserting (\ref{fin}) into
(\ref{introlast}) and contracting with $\5g^{ik}\6_k r$ we get
\be
\7E^i\6_i r = {\Delta_5 \over (r\sin\t)'}\,\5g^{ij}\6_i r
\6_j (r\cos\t)
\label{fin2}
\ee
which, when inserted back into (\ref{introlast}), allows us to find
the following relation between $r$ and $\tilde E^i$:
\be
\left(r\sin\t\right)'\7E^i =
\Delta_5 \5g^{ij}\left[\left(1+{\5g^{kl}\6_k r\6_l r \over r^2}\right)
\6_j(r\cos\t) -\6_j r\,
{\5g^{kl}\6_k r \6_l (r\cos\t)\over r^2} \right]\, .
\label{bpsham}
\ee
Then, using the identities
\bea
\left(r\sin\t\right)' &=&
\left(1+{\5g^{kl}\6_k r\6_l r \over r^2}\right) r cos\t
-{\5g^{kl}\6_k r \6_l (r\cos\t) \over r} \\
\5g^{\8{\i}\8{\j}}\6_{\8{\i}}r\6_{\8{\j}}r\sin\t  & = &
-{\5g^{kl}\6_k r\6_l r \over r}\left(r\cos\t\right)'
 + r'\, {\5g^{kl}\6_k r \6_l (r\cos\t)\over r}\, ,
\eea
one can check that (\ref{bpsham}) is equivalent to the two BPS equations
(\ref{BPSe1}) and (\ref{BPSe2}). When combined with the Gauss law these are
equivalent to the second order equation (\ref{pde}). We conclude that
configurations satisfying this equation preserve 1/4 supersymmetry.

\subsection{Multi-angle D5-brane solutions with 1/4 supersymmetry}

We have just shown that any solution $r(\theta,\theta_{\8{\i}})$
to equation (\ref{pde}) preserves $1/4$ supersymmetry. We earlier discussed
the $SO(5)$ invariant solution to this equation. We recall here that it has
the form \cite{callan}
\begin{equation}
r_{\nu}(\theta)=C{\left(\eta(\theta)\right)^{1/3}\over \sin\theta} \, ,
\label{invsol}
\end{equation}
We shall now show that there 
exist many other, less-symmetric, solutions, in the
near horizon region. Some may be found by considering fluctuations about the
$SO(5)$-invariant solution. This analysis suggests that a large class of
solutions may be found via an ans{\"a}tze of the form 
\begin{equation}
r(\theta,\theta_{\8{\i}})=r_{\nu}(\theta)\prod_{\8{\i}}\delta_{\8{\i}}
(\theta_{\8{\i}})\, .
\label{ans}
\end{equation}
When this is inserted into eq. (\ref{pde}) we find that
\begin{eqnarray}\label{f1}
0 & = & \sin^2\theta\sin^2\theta_3\sin\theta_2\left[3\sin^2\theta_4\cos\theta_4
\psi_4 + \sin^3\theta_4\psi_4^\prime\right] \nonumber \\
& & +\sin^2\theta\sin\theta_4\sin\theta_2\left[2\sin\theta_3\cos\theta_3
\psi_3 + \sin^2\theta_3\psi_3^\prime\right] \nonumber \\
& & + \sin^2\theta\sin\theta_4\left[\cos\theta_2\psi_2 + \sin\theta_2
\psi_2^\prime\right] + {\sin^2\theta\sin\theta_4\over
\sin\theta_2}\psi_1^\prime \nonumber \\
& & -\sin^3\theta_4\sin^2\theta_3\sin\theta_2{d\over d\theta}\left({r_\nu
\sin^3\theta \over (r_\nu\sin\theta)^\prime}\right)\left(\psi_4^2 + {\psi_3^2
\over \sin^2\theta_4} \right. \nonumber \\
& & \left. + {\psi_2^2\over \sin^2\theta_4\sin^2\theta_3}
+ {\psi_1^2\over \sin^2\theta_4\sin^2\theta_3\sin^2\theta_2}\right) \, ,
\end{eqnarray}
where the new functions $\psi_{\8{\i}}(\theta_{\8{\i}})$, each depending on 
just one angular variable $\theta_{\8{\i}}$, are defined by
\begin{equation}
\psi_{\8{\i}}(\theta_{\8{\i}}) \equiv
{\delta_{\8{\i}}^\prime(\theta_{\8{\i}})\over
\delta_{\8{\i}}(\theta_{\8{\i}})}\, ,
\end{equation}
the prime indicating differentiation with respect to the corresponding
angular variable.
The $SO(5)$ invariant solution (\ref{invsol}) satisfies the identity
\begin{equation}
{d\over d\theta}\left({r_{\nu}(\theta)\sin^3\theta\over \left(r_{\nu}(\theta)
\sin\theta\right)^\prime}\right)=3\sin^2\theta \quad ,
\label{inveq}
\end{equation}
which can be used to simplify (\ref{f1}) to
\footnote{In fact, by requiring ${d\over d\theta}\left({r_{q,\nu}(\theta)
\sin^3\theta\over \left(r_{q,\nu}(\theta)
\sin\theta\right)^\prime}\right)=(3+q)\sin^2\theta$, we can obtain more
general ans\"{a}tze compatible with the separation of variables whose
dependence on the polar angle $r_{q,\nu}(\theta)$ is different from the one 
in the $SO(5)$ invariant solution (\ref{invsol}). We will not consider these 
solutions here.} 
\begin{eqnarray}
0 & = & \sin^2\theta_3\sin\theta_2\left[3\sin^2\theta_4\cos\theta_4
\psi_4 + \sin^3\theta_4\psi_4^\prime - 3\sin^3\theta_4\psi_4^2\right]
\nonumber \\
& & + \sin\theta_4\sin\theta_2\left[2\sin\theta_3\cos\theta_3\psi_3 +
\sin^2\theta_3\psi_3^\prime -3\sin^2\theta_3\psi_3^2\right]
\nonumber \\
& & + \sin\theta_4\left[\cos\theta_2\psi_2 + \sin\theta_2
\psi_2^\prime - 3\sin\theta_2\psi_2^2\right] \nonumber \\
& & + {\sin\theta_4\over \sin\theta_2}\left[\psi_1^\prime-3\psi_1^2\right]
\label{f2}
\end{eqnarray}
A simple way to find solutions of this equation is to require  
$\psi_{\8{\i}}$ to satisfy the Bernoulli equations
\footnote{This is certainly not the most general solution to eq. (\ref{f2}).
One may introduce three separation constants following the usual separation of 
variables prescription.}
\begin{eqnarray}
\psi_4' + 3\cot\t_4 \,\psi_4 & = & 3\psi^2_4
\label{b1} \\
\psi_3' + 2\cot\t_3 \,\psi_3 & = & 3\psi^2_3
\label{b2} \\
\psi_2' + \cot\t_2 \,\psi_2 & = & 3\psi^2_2
\label{b3} \\
\psi_1' & = & 3\psi^2_1\, .
\label{b4}
\end{eqnarray}
These equations admit the trivial solutions $\psi_{\8{\i}}(\theta_{\8{\i}})=0$,
whereby we recover the previous $SO(5)$ invariant solution, but they are also
solved by
\begin{eqnarray}
{1\over \psi_4} & = & c_4\sin^3\t_4 + {3\over 2}\sin\t_4\cos\t_4 -{3\over 2}
\sin^3\t_4\log\,\tan {\t_4\over 2} \label{s1} \\
{1\over \psi_3} & = & c_3\sin^2\t_3 + 3\sin\t_3\cos\t_3 \label{s2} \\
{1\over \psi_2} & = & c_2\sin\t_2 -3\sin\t_2\log\,\tan{\t_2\over 2}
\label{s3} \\
{1\over \psi_1} & = &  c_1 -3\t_1 \label{s4}
\end{eqnarray}
where $c_{\8{\i}}$ are constants of integration.

Due to the functional form of our ansatz (\ref{ans}), the new solutions
cannot represent isolated singularities. In fact, the set of singularities
in the radial function $r(\theta,\theta_{\8{\i}})$ is the union of the
corresponding sets of singularities of the different product terms
appearing in (\ref{ans}). Let us analyse the singularity structure
of the new functions $\delta_{\8{\i}}(\theta_{\8{\i}})$. They are formally
given by
\begin{equation}
\delta_{\8{\i}}(\theta_{\8{\i}})=a_{\hat {\i}}e^{\int \psi_{\hat {\i}}
(\theta_{\hat{\i}})\, d\theta_{\hat {\i}}} \, ,
\end{equation}
which can not be solved analytically, in general. It is nevertheless
possible to study their behaviour near singular points. Given a Bernoulli
equation of the type :
\begin{equation}\label{geq}
\psi'\,+\,q\,\cot\theta\,\psi\,=\,3\,\psi^2 \quad , \quad q=1,2,3
\end{equation}
it can be checked that,
\begin{eqnarray}
& \psi\sim
\cases{
{q-1\over 3}\,{1\over \theta}& if $q\not= 1$\cr
-{1\over 3}\,{1\over \theta\log\theta}& if $q= 1$\cr}
\,\,\,\,\,\,\,\,\,\,\,\,\,\,\,\,\,\,\,\,\,\,\,\,
(\,\theta\sim\,0\,) & \label{asy1} \\\\
& \psi\sim
\cases{
{q-1\over 3}\,{1\over \theta-\pi}&if $q\not= 1$\cr
-{1\over 3}\,{1\over (\theta-\pi)\log(\pi-\theta)}& if $q= 1$\cr}
\,\,\,\,\,\,\,\,\,\,\,\,\,\,\,\,\,\,\,\,\,\,\,\,
(\,\theta\sim\,\pi\,) & \label{asy2}
\end{eqnarray}
from which one can conclude that new functions $\delta$ are non-singular
around $\theta\sim 0,\pi$. Since the function $1/\psi$ changes sign
when $\theta$ runs from $0$ to $\pi$, we conclude that it should
vanish for some value $\theta=\alpha$ defined by
\begin{equation}\label{sing}
{1\over \psi(\alpha)} = 0 \quad , \quad \alpha\neq 0,\pi
\end{equation}
Since (\ref{geq}) can be rewritten for $\psi\neq 0$ as
\beq
-{d\over d\theta}\,\Big(\,{1\over \psi}\,\Big)\,+\,
q\cot\theta \,{1\over \psi}\,=\, 3 \, ,
\eeq
it is straightforward to derive the behaviour
\beq
\psi\sim-{1\over 3}\,{1\over \theta-\alpha}
\,\,\,\,\,\,\,\,\,\,\,\,\,\,\,\,\,\,\,\,\,\,\,\,
(\,\theta\sim\,\alpha\,)
\eeq
which is independent of $q$. Thus all non-polar angles have branch
points of the type
\beq
\delta\sim{1\over (\theta-\alpha)^{{1\over 3}}} \, .
\eeq

The analysis for $q=0$, which applies to the angle $\theta_1$,  is
straightforward. In this case $\delta_1(\theta_1)$
can be computed analytically:
\begin{equation}\label{anal}
\delta_1(\theta_1)={a_1\over(\alpha_1-\theta_1)^{1/3}}\, .
\end{equation}
Note that the reality of $r(\theta,\theta_{\8{\i}})$ restricts $\theta_1$ to the
interval $(0,\alpha_1)$, so we have a case of partial wrapping\footnote{We may
assume that $\alpha \leq 2\pi$ since we may otherwise shift $\theta_1$ by
$2\pi$.}. That is, the branch cut at $\theta_1=\alpha_1$ represents a locus of
points at which the D5-brane leaves the 5-sphere. These configurations are
therefore not `baryonic branes' as defined in the introduction, although they
are 1/4 supersymmetric configurations of the D5-brane. 

Inserting the ansatz (\ref{ans}) into equations (\ref{BPSe1})-(\ref{BPSe2}),
we can compute the conjugate momentum $E^i$ for any solution of the system
(\ref{b1})-(\ref{b4}):

\begin{eqnarray}
E^4 & = & R^4\sin^2\t\,\sqrt{det\,^4g}\,\psi_4(\t_4)
\label{cm1} \\
E^3 & = & R^4\sin^2\t\,\sqrt{det\,^4g}\,{\psi_3(\t_3)\over \sin^2\t_4}
\label{cm2} \\
E^2 & = & R^4\sin^2\t\,\sqrt{det\,^4g}\,{\psi_2(\t_2)\over
\sin^2\t_4\sin^2\t_3}
\label{cm3} \\
E^1 & = & R^4\sin^2\t\,\sqrt{det\,^4g}\,{\psi_1(\t_1)\over
\sin^2\t_4\sin^2\t_3\sin^2\t_2}
\label{cm4} \\
E^\t & = & E_\nu(\t)\,\sqrt{det\,^4g} - {3\over 2}R^4\eta(\t)\sqrt{det\,^4g}
\left[(\psi_4(\t_4))^2 \right. \nonumber \\
& & \left. + {(\psi_3(\t_3))^2 \over \sin^2\t_4} +{(\psi_2(\t_2))^2 \over
\sin^2\t_4\sin^2\t_3} + {(\psi_1(\t_1))^2\over 
\sin^2\t_4\sin^2\t_3\sin^2\t_2}\right]\, .
\label{cm5}
\end{eqnarray}

A simple calculation suffices to show that the $E^i$ satisfy the Gauss law
constraint (\ref{gausslaw}) as a consequence of the differential equations
satisfied by the functions $\psi_{\hat{\i}}$. To compute the energy of the new
configurations, we should compute their topological charge. Using 
the ansatz
(\ref{ans}) in (\ref{topo}), this can be written as

\begin{eqnarray}
{\cal Z}_5 & = & \sqrt{det\,^4g}\prod_{\hat {\i}}\delta_{\hat {\i}}
(\t_{\hat {\i}})\left({\cal Z}(\t) + {3\over 2}r_0(\t)
{\eta(\t)\over \sin\t}R^4\left[(\psi_4(\t_4))^2 \right.
\right. \nonumber \\
& & \left.\left. + {(\psi_3(\t_3))^2\over \sin^2\t_4} +
{(\psi_2(\t_2))^2\over \sin^2\t_4\sin^2\t_3} + {(\psi_1(\t_1))^2\over 
\sin^2\t_4\sin^2\t_3\sin^2\t_2}
\right]\right)
\label{tc}
\end{eqnarray}
The terms associated with
new ($\psi_{\8{\i}}$-dependent) contributions  are such that
the new solutions have a positively defined topological charge whenever the
`old' ($SO(5)$-invariant) contribution is non-negative. This is the case, for
the full range of $\theta$, when $\nu=0$. As expected from our general
discussion,
${\cal Z}_5$ can be rewritten as a divergence, 
${\cal Z}_5=\partial_i {\cal Z}^i_5$, at least in the near-horizon
($a=0$) limit.  By
inserting the conjugate momenta (\ref{cm1}-\ref{cm5}) into 
(\ref{div1}), taking the near horizon limit and using the 
ans\"atze (\ref{ans}) for our new radial functions, we find the components
$\vec{{\cal Z}}_5$ to be
\bea
{\cal Z}_5^\t & = & E^\t\,r\cos\t +R^4\,\sin^5\t\,\sqrt{det\,^4g}\,r
\nonumber \\
{\cal Z}_5^{\8{\i}} & = & \,E^{\8{\i}}\,r\cos\t \, .
\label{div11}
\eea 

To summarize, we have found many new 1/4 supersymmetric configurations of a
D5-brane in the near-horizon geometry of a D3-brane background. None of these
new solutions has a straightforward 
interpretation as a `baryonic brane' because
(i) they have branch cut singularities rather than poles, and (ii) they only
partially wrap the 5-sphere. However, it seems clear that there must be a
large class of other 1/4 supersymmetric solutions that are not captured by our
ansatz. We have argued earlier that these are likely to include solutions with
multiple isolated point singularities, but it appears to be difficult to find
explicit solutions of this type.

\setcounter{equation}{0}
\section{Baryonic M5-brane}

We start from the M5-brane solution of D=11 supergravity. The metric and
4-form field strength $F_{(4)}=dC^{(3)}$ are
\bea
ds^2_{(11)} & = & U^{-1/3}\left[ds^2(\bE^{(1,4)}) + dX^2_5\right] +
U^{2/3}\left[dr^2 + r^2\,d\Omega^2_{(4)}\right] \\
F_{(4)} & = & -3\,R^3 \omega_4
\eea
where $d\Omega_4^2$ is the $SO(5)$ invariant metric on the unit 4-sphere
and $\omega_4$ is its volume 4-form.
The function $U$ is
\be
U= a + \left({R\over r}\right)^3\quad , \quad
\left(R^3=\pi\,Nl^3_p\right)\, .
\label{harmonicM}
\ee

We will put a probe M5-brane in this background and look for solutions
for which the M5-brane wraps around the $S^4$ factor. We shall start from the
Lorentz covariant M5-brane action \cite{sorokin}; a useful review is
\cite{sorokinb} to which we refer for details of this formalism and references
to related work. We use the conventions of \cite{BST}, with minor
modifications.
The action is constructed from the induced worldvolume metric $h$ and a 2-form
worldvolume potential $A$ with `modified' 3-form field strength
\be
H=dA - C^{(3)}\, .
\ee
Here, $C^{(3)}$ should be understood as the pullback of the 3-form potential of
the background; we will use the same symbol for a spacetime form and its
pullback since it will be clear which is meant from the context. Thus, the
Bianchi identity for $H$ can be written as
\be\label{Mbianchi}
dH = -F_{(4)}\, .
\ee
There is an additional field in the Lorentz covariant action, the `PST scalar',
but it can be removed by a gauge transformation at the cost of
breaking manifest
Lorentz covariance.

Let $\xi^\mu= (t,\sigma^I)$ be the M5-brane worldvolume coordinates, and define
the {\sl worldspace} tensor
\be
\tilde H^{IJ}= {1\over 6\sqrt{\det g}} \varepsilon^{IJKLMN} H_{KLM}
\ee
where $g$ is the induced {\sl worldspace} metric. Let
\be
\tilde H_{\mu\nu} = h_{\mu I}h_{\nu J} \tilde H^{IJ}
\ee
This is not a worldvolume tensor but it is derived from a tensor by choice of
the temporal gauge for the PST field. In the same gauge, and for unit M5-brane
tension, the M5-brane action is
\be\label{22}
S = -\int\! d^6\xi\,\left\{ \sqrt{-\det(h+ \tilde H)} +
V^Ph_{P0}\right\} + {1\over4}\int \left[C^{(6)} + 2 H\wedge C^{(3)}
+ H\wedge H \right]
\ee
where
\be
V^P = {1\over 24} g^{PN}\varepsilon^{IJKLM}H_{KLM} H_{IJN}
\ee
and $C^{(6)}$ is the (pullback of) the 6-form potential dual to $C^{(3)}$
(defined in backgrounds that satisfy the D=11 supergravity field
equations).

Notice that in the given background, the Bianchi identity (\ref{Mbianchi})
becomes
\be
dH=3\,R^3\omega_4 \,.
\label{bity}
\ee
As in the D5-brane case we shall find that this will require infinite radial
deformations carrying M2-brane charge, analogous to the self-dual string
solitons
on the M5-brane \cite{HLW}. Pursuing this analogy, we may think of the entire
configuration as one involving three branes, background, probe, and `soliton',
intersecting according to the array
\be
\ba{cccccccccccl}
M5: &1&2&3&4&5&\_&\_&\_&\_&\_ &\quad \mbox{background}    \nonumber \\
M5: &\_&\_&\_&\_&5&6&7&8&9&\_ &\quad \mbox{probe}     \nonumber \\
M2: &\_&\_&\_&\_&5&\_&\_&\_&\_&\natural &\quad \mbox{soliton.}
\ea
\label{dualtriplea}
\ee
The symbol $\natural$ is read as `ten'. The tenth direction is the radial
direction.

The probe M5-brane has bosonic worldvolume fields $X^0$, $X^a$
($a=1,2,3,4$), $X^5$, $r$ and $\T^i$ where $\T^i$ ($i=1,\dots,4)$ are now four
angles parametrising $S^4$, and a three-form field strength $H$. Let
$\sigma^I = (\sigma,\theta^i)$ be the worldspace coordinates with $\t^i$
parameterising the 4-sphere. We shall choose the static gauge
\be
X^0=t\,\qquad X^5=\sigma\, , \qquad  \T^i = \theta^i\qquad (i=1,2,3,4)\, ,
\ee
appropriate to an M5-brane wrapped on a 4-sphere at a (variable) radius in
the 5-space transverse to the background M5-brane. We also set
\be
X^1 =X^2=X^3=X^4= 0\, .
\ee
In other words, we assume that $r$ is the only `active' scalar. We
also assume that $\partial_5 r=0$, so that $r$ is a function on the
4-sphere, with isolated singularities. For static configurations we now have
\be
h_{\mu\nu} = \pmatrix{-U^{-1/3} & 0 & 0\cr 0& U^{-1/3} & 0\cr
0&0& g_{ij}}
\ee
where
\be\label{Mgee}
g_{ij}=U^{2\over3}\left(r^2\5g_{ij} + \6_i r\6_j r\right)\, .
\ee
The metric $\bar g$ is now the $SO(4)$-invariant metric on the unit
3-sphere.

For the background we are considering, $C^{(3)}_{5ij}$ vanishes, so we
may assume that
\be
H_{5ij}=0\, .
\ee
This means that the worldspace vector density $V^I$ vanishes and that the only
non-vanishing components of $\tilde H^{IJ}$ are
\be
\tilde H^{5i}= {U^{1\over6}\over \sqrt{\det g}}\, \Pi^i
\ee
where
\be
\Pi^i \equiv {1\over 6} \varepsilon^{ijkl} H_{jkl}\, .
\ee
In terms of this new variable the Bianchi identity (\ref{bity}) is
\be
\partial_i \Pi^i = 3\,R^3\,\sqrt{\det\5g}\, .
\label{bity1}
\ee
Since $g_{5i}=0$, the only non-vanishing components of $\7H_{\mu\nu}$ are
\be
\7H_{5i}=g_{55}g_{ij}\7H^{5j}={U^{-1/6}\over \sqrt{\det g}}\, g_{ij}\Pi^j\, .
\ee

\subsection{Energy bound}

We shall now derive an energy bound for baryonic M5-branes.
It follows from the Hamiltonian formulation of the super M5-brane
in a general background \cite{BST}, that the energy density ${\cal H}$ of a
static configuration, in the background we consider and subject to the
restrictions discussed above, is such that
\be
{\cal H}^2 = U^{-{2\over 3}}\left[\det g + \Pi^i\Pi^j g_{ij}\right]\,
\label{ed}
\ee
where $g_{ij}$ is given by (\ref{Mgee}).
This expression is invariant under the $SO(5)$ isometry group of the
4-sphere, but minimum energy configurations can be at most $SO(4)$
invariant. We shall later wish to exploit this possibility so we write
the 4-sphere metric as
\be
ds^2 = d\theta^2 + \sin^2\theta\, d\Omega_3^2
\ee
where $d\Omega_3^2$ is the $SO(4)$-invariant metric on the unit 3-sphere.
Let $\theta^{\8{\i}}$ ($\8{\i}=1,2,3$) parametrize the 3-sphere,
so that $\t^i=(\t,\t^{\8{\i}})$. We may now rewrite (\ref{ed}) as
\bea
{\cal H}^2 & = & \Delta_4^2 \left(r^2 + r'^2 + \5g^{\8{\i}\8{\j}}\6_{\8{\i}} r
\6_{\8{\j}} r\right) + \left(\Pi^\t r'\right)^2 +
\left(\Pi^{\8{\i}}\6_{\8{\i}}r\right)^2 \nonumber \\
& & + 2(\Pi^\t r')(\Pi^{\8{\i}}\6_{\8{\i}}r) + \left(\Pi^\t r\right)^2 +
r^2 \Pi^{\8{\i}} \Pi^{\8{\j}}\5g_{\8{\i}\8{\j}}\, ,
\label{in1}
\eea
where
\be
\Delta_4 \equiv U\,r^3\,\sqrt{\det\5g} \, .
\label{redef}
\ee
This can be further rewritten as
\be
{\cal H}^2 = {\cal Z}_4^2 + \left[\Delta_4\left(r\cos\t\right)' +  \Pi^i\6_i
\left(r\sin\t\right)
\right]^2 + |\Delta_4\5g^{\8{\i}\8{\j}}\6_{\8{\j}}r + r\,\Pi^{\8{\i}}|^2
\label{Men}
\ee
where
\be
{\cal Z}_4 = \Delta_4 \left(r\sin\t\right)' - \Pi^i\6_i \left(r\cos\t
\right)
\ee
and $||^2$ indicates contraction with the $\8{\i}\8{\j}$ components of the
four sphere metric $\bar g_{ij}$, and $\5g^{\8{\i}\8{\j}}$ are the
$\8{\i}\8{\j}$ components of the inverse 4-sphere metric $\5g^{ij}$.

Using the  Bianchi identity (\ref{bity}), one can show that ${\cal Z}_4= \6_i
{\cal Z}_4^i$ where
\bea
{\cal Z}_4^\t & = & -\Pi^\t\,r\cos\t +\sqrt{\det\5g}\,\sin\t
\left(a\,\,{r^4\over 4} + r\,R^3\right) \nonumber \\
{\cal Z}_4^{\8{\i}} & = & -\,\Pi^{\8{\i}}\,r\cos\t\, .
\label{div2}
\eea
{}From (\ref{Men}), and the fact that ${\cal Z}_4$ is a divergence, we
deduce the bound
\be\label{Mb}
{\cal H}\geq |{\cal Z}_4|
\ee
with equality when
\bea
\Pi^{\8{\i}} & = & -\,\Delta_4 {\5g^{\8{\i}\8{\j}}\6_{\8{\j}}r \over r}
\label{bpse1} \\
\Pi^\t & = & -\,{\Delta_4 \over \left(r\sin\t\right)'}
\left(\left(r\cos\t\right)' - \5g^{\8{\i}\8{\j}}\6_{\8{\i}}r\6_{\8{\j}}r
{\sin\t\over r}\right) \quad .
\label{bpse2}
\eea

Because ${\cal H}$ is, by assumption, independent of
the worldspace coordinate
$\sigma$ we may interpret its integral over the remaining worldspace
coordinates, parameterising $S^4$, as the tension $T$ of a string with string
coordinate $\sigma$. It then follows from the bound (\ref{Mb}) that
\be
T\ge \int\! d^4\theta\, \big|{\cal Z}_4\big|
\ge \bigg|\int\! d^4\theta {\cal Z}_4\,\bigg|\, .
\label{tensionT}
\ee
The first inequality is saturated by solutions of (\ref{eqdiff1}). The second
inequality will be saturated too provided that ${\cal Z}_4$ does not
change sign
in the integration region. If this is the case then the tension $T$ can be
expressed as a surface integral over a sum of $N$ 3-spheres surrounding the N
singular points on the 4-sphere. Each such surface term can be considered to be
the energy/charge of an M2-brane attached to the 4-sphere at that point.

By combining the BPS conditions (\ref{bpse1}) and (\ref{bpse2}) with the
Bianchi
identity (\ref{bity1}) we deduce that the energy density is minimised when the
radial function $r$ is a solution of the second order equation
\be
\left[{\Delta_4 \over \left(r\sin\t\right)'}
\left(\left(r\cos\t\right)' - {\5g^{\8{\i}\8{\j}}\6_{\8{\i}}r\6_{\8{\j}}
\left(r\sin\t\right)\over r}\right)\right]' +
\6_{\8{\i}}\left[\Delta_4 {\5g^{\8{\i}\8{\j}}
\6_{\8{\j}}r \over r}\right]
=-3\,R^3\,\sqrt{\det\5g}\, .
\label{eqdiff1}
\ee

In the remainder of this subsection we shall restrict ourselves to
the near-horizon $a=0$ geometry.
To simplify the task of finding explicit solutions of  equations
(\ref{bpse1}) and (\ref{bpse2}) in this case, we will
restrict our attention to an $SO(4)$-invariant ansatz for which the radial
function depends only on $\theta$. In this case only the $\Pi^\t$ component of
$\Pi^i$ is non-zero, and
\be
\Pi^\t = \sqrt{\det\5g^{(3)}}\,\Pi(\t)
\ee
where $\Pi(\t)$ is a function only of $\t$, and $\5g^{(3)}$ is the
$SO(4)$-invariant metric on the unit 3-sphere. The Bianchi identity
(\ref{bity}) now reduces to
\be
\Pi'=3\,R^3\,\sin^3\t \, .
\label{gl}
\ee
This has the solution
\be
\Pi(\theta)\,=-\,R^3\,\left[\cos\t\left(\sin^2\t +
2\right)+2(2\nu-1)\right]\, ,
\label{eqpi}
\ee
where $\nu$ is an integration constant which, as in the D5-brane case, can be
restricted to lie in the interval $[0,1]$.

It will be convenient to set
\be
\Delta_4 = \sqrt{\det \bar g^{(3)}}\, \Delta(\t)
\ee
since $\Delta = R^3 \sin^3\t$ is a function only of $\t$. For
$SO(4)$-invariant M5-brane configurations the first BPS equation (\ref{bpse1})
is trivially satisfied while (\ref{bpse2}) reduces to
\be\label{redbps}
{r'\over r}={\Delta\sin\t -\Pi\cos\t \over \Delta\cos\t +\Pi\sin\t}\, .
\ee
We may also set
\be
{\cal Z}_4 = \sqrt{\det \bar g^{(3)}}\,{\cal Z}
\ee
where ${\cal Z}$ is now a function only of $\t$. Using (\ref{redbps}) we find
that
\be
{\cal Z} = r\, {\left(\Delta \sin\t -\Pi\cos\t\right)^2 +\left(\Delta\cos\t
+\Pi\sin\t\right)^2 \over \left(\Delta\cos\t + \Pi\sin\t\right)}\, .
\label{signMbrane}
\ee
Thus ${\cal Z}$ has a definite sign provided that the denominator does not
vanish. As
\be
\Delta\cos\t + \Pi\sin\t = -2R^3\sin\t\left[\cos\t -1+2\nu\right]\, ,
\ee
the denominator of (\ref{signMbrane}) has a definite sign for $\t\in
[0,\pi]$ if $\nu=0$. In this case the solution of the first-order differential
equation (\ref{redbps}) can be easily shown to be:
\beq
r\,=\,r_0 \, {\rm sec} (\theta/2)
\label{solM5}
\eeq
where $r_0$ is the value of $r(\t)$ at $\theta=0$. This solution
was previously found in \cite{callan2} in the context of
a D4-brane probe in the D4-brane
background geometry. It is straightforward to verify that the function
given in (\ref{solM5}) solves the second-order ODE obtained from
(\ref{eqdiff1}) by setting $a=0$ and restricting $r$ to depend only
on $\theta$, namely:
\beq
\left[\,\sin^3\t\,{r'\cos\t-r\sin\t \over
r'\sin\t + r\cos\t}\right]' = -3\,\,\sin^3\t \, .
\label{ed1}
\eeq
The function (\ref{solM5}) represents an M5-brane wrapped on the $S^4$ with a
spike at the point $\theta=\pi$, where $r(\theta)$ diverges. This spike
extends along the axis $\theta=\pi$. From our previous
results it is a simple exercise to compute the value of the tension $T$ for the
configuration represented by (\ref{solM5}). Indeed, if we recall that
${\cal Z}_4$ is a total derivative, the integral of
${\cal H}$ over the four-sphere can be immediately evaluated. The energy per
unit surface of the spike obtained from this calculation is:
\beq
T_{M5}\,V_{(3)}\,\,|\,\Pi\,(\,\pi\,)|\,\,,
\eeq
where $V_{(3)}\,=\,2\pi^2$ is the volume of the unit three sphere,
$T_{M5}$ is the tension of  the M5-brane which, up to now,  was considered
to be
one and which is given by:
\beq
T_{M5}\,=\,{1\over (2\pi)^5\,l_p^6}\,\,,
\eeq
and $\Pi\,(\,\pi\,)$ can be extracted from
(\ref{eqpi}) with $\nu=0$. An elementary  calculation shows that
\be
T_{M5}\,V_{(3)}\,\,|\, \Pi\,(\,\pi\,)|\,=\,
\,N\,T_{M2}\, ,
\label{tension}
\ee
where
\be
T_{M2}\,=\,{1\over (2\pi)^2\,l_p^3}
\ee
is the tension of the M2-brane. This clearly shows
that the spike at $\theta=\pi$ can be interpreted as a bundle of
$\,N$ M2-branes emanating from the M5-brane. The relation of our
configuration of M5-branes and M2-branes with a configuration of D4-branes
and IIA fundamental strings is easily understood if we reduce the array
(\ref{dualtriplea}) along the 5-direction.

As in the D5-brane case, we could have considered configurations with
$\nu\not=0$. The corresponding solutions of the BPS equation can be extracted
from those found in \cite{callan2} for the D4-brane in the near-horizon region
of the D4-brane geometry. These configurations of the M5-brane do not
completely wrap the $S^4$ and they extend into the M5-brane `throat' region
all the way to $r=0$, in complete analogy with what happens in the D5-brane
case. Similarly, one can evaluate the energy per unit surface of the spikes of
the $\nu\not=0$ solutions and show that their energy per unit surface
corresponds to a bundle of $k$ M2-branes with $k<N$.

\subsection{Full M5 background}

We are now going to study the embedding of the probe M5-brane in the full
M5-brane metric, which corresponds to taking $a=1$ in the harmonic function
$U$
of equation (\ref{harmonicM}). We shall restrict
ourselves to the analysis of the $SO(4)$-invariant embeddings. In this
case, as was shown in \cite{callan}, it is more convenient to work in a
new set of variables $(z,\rho)$, related to $(r,\theta)$ as follows:
\beq
z\,=\,-r\,{\rm cos }\,\theta\,\,,
\,\,\,\,\,\,\,\,\,\,\,\,\,\,\,\,\,\,\,\,\,\,\,\,\,\,\,\,
\rho\,=\,r\,{\rm sin }\,\theta\,\,.
\label{extraone}
\eeq
It is clear from (\ref{extraone}) that $z$ can take values in the interval
$(-\infty, +\infty)$ whereas $0\le\rho< \infty$.
In these new coordinates the M5-brane embedding is determined by a function
$z(\rho)$. By using the relation between the $(z,\rho)$ and $(r,\theta)$
coordinates, it is not difficult to relate  $z(\rho)$ and its derivative to
$r(\theta)$ and $r'(\theta)=dr/d\theta$. After a short calculation one gets:
\beq
{dz\over d\rho}\,=\,
{r\,{\rm sin }\,\theta\,-\,{\rm cos }\,\theta\,\,\,r\,'\over
r\,{\rm cos }\,\theta\,+\,{\rm sin }\,\theta\,\,\,r\,'}\,\,.
\label{extratwo}
\eeq
By using (\ref{extratwo}) one can convert the BPS condition
(\ref{bpse2}) into a first-order differential equation for $z(\rho)$. To this
end, we introduce the function
\beq
\hat\Delta(\rho,z)\,=\,\rho^{3}\,\,
\Big(\,1\,+\,{R^{3}\over [\,\rho^2\,+\,z^2]^{{3\over 2}}}
\,\,\Big)
\label{extrathree}
\eeq
It is easy to show that (\ref{bpse2}) is now equivalent to
\beq
{dz\over d\rho}=\,
{\Pi\Bigr(\,{\rm arctan}\,(\,-\rho/z)\,\Bigr)\over
\hat\Delta(\rho,z)}\, ,
\label{extrafour}
\eeq
where $\Pi$ is the same function as in equation (\ref{eqpi}) and its argument
in (\ref{extrafour}) is obtained after inverting the relation (\ref{extraone})
between both coordinate systems.

The condition (\ref{extrafour}) can also be obtained by studying the energy of
the M5-brane in the new variables. Let us define, as in equation
(\ref{tensionT}), the tension $T$ as the energy density integrated over the
four worldspace coordinates $\theta^i$ ($i=1,\cdots,4$). It is not difficult to
change variables in this integral and express the result as the following
integral over $\rho$:
\beq
T\,=\,V_{(3)}\,
\int d\rho\,
\sqrt{\bigl(\,\Pi\,{dz\over d\rho}\,+\,\hat\Delta\,\bigr)^2\,+\,
\bigl(\,\hat\Delta\,{dz\over d\rho}\,-\,\Pi\,\bigr)^2}
\label{extrafive}
\eeq
In this equation, and in what follows, $\Pi$ depends on $\rho$ and $z(\rho)$ as
in (\ref{extrafour}). The right-hand side of equation (\ref{extrafive})
contains a remarkably simple sum of squares. This simplicity is a reflection of
how convenient are the new coordinates $(z,\rho)$ for this $a=1$ case, as will
be confirmed below. Notice that the second term in
(\ref{extrafive}) vanishes precisely when the BPS condition (\ref{extrafour})
holds. Clearly, if we define:
\beq
X\,\equiv\,\Pi\,{dz\over d\rho}\,+\,\hat\Delta
\label{extrasix}
\eeq
we obtain the following bounds:
\beq
T\,\ge\,V_{(3)}\,\int d\rho\,\big|\,X\,\big|\,\ge\,V_{(3)}\,
\bigg|\,\int d\rho\,X\,\bigg|\,\,.
\label{extraseven}
\eeq
Obviously, the first bound in (\ref{extraseven}) is saturated when the BPS
condition (\ref{extrafour}) holds. We will show below that the second
inequality is also saturated if equation (\ref{extrafour}) is satisfied.
Moreover, it is easy to prove that $X$ is a total derivative with respect to
$\rho$, namely:
\beq
X\,=\,{d\over d\rho}\,
\Bigl[\,z\Pi\,+\,
\Bigl(\,{1\over 4}\,+\,{R^3\over [\,\rho^2\,+\,z^2]^{{3\over 2}}}\,\Bigr)
\,\rho^4\,\Bigr]\,\,.
\label{extraeight}
\eeq
It is important to stress the fact that equation (\ref{extraeight}) follows
from the explicit form of $\Pi$ given in (\ref{eqpi}) and, thus, is valid for
any function $z(\rho)$. Actually, the derivative of $\Pi$ with respect to
$\rho$ for an arbitrary function $z(\rho)$ is given by:
\beq
{d\Pi\over d\rho}\,=\,
{3R^3\rho^3\over [\,\rho^2\,+\,z^2]^{{5\over 2}}}\,\,
\bigl(\,\rho\,{d z\over d\rho}\,-\,z\,\bigr)\,\,.
\label{extranine}
\eeq
Using  (\ref{extranine}) to evaluate the right-hand side of
(\ref{extraeight}), one easily verifies that the result is the function
defined in equation (\ref{extrasix}). Moreover, the value of $X$ when the BPS
condition is satisfied is given by:
\beq
X_{BPS}\,=\,\Big[\,{dz\over d\rho}\,+\,
\Big({dz\over d\rho}\Bigr)^{-1}\Big]\,\Pi\,\,.
\label{extraten}
\eeq
As $\hat\Delta\ge 0$, the sign of $dz/d\rho$ for any  $z(\rho)$
satisfying (\ref{extrafour}) is the same as the sign of $\Pi$. Therefore, it
follows that
\beq
X_{BPS}\,\ge\,0
\label{extraeleven}
\eeq
which  implies, as was previously claimed,  that the second inequality in
(\ref{extraseven}) is saturated if the first one is saturated.

The BPS equation (\ref{extrafour}) is the same as the one appearing in
the study of a D4-brane probe in the full D4-brane geometry. Its analytical
solution was found in \cite{ramallo}. It turns out that $z(\rho)$ can be
represented implicitly by the equation
\beq
z\,=\,z_{\infty}\,+\,{R^3\over \rho^2}\,\,
\Bigl[\,\,{z\over \sqrt{\rho^2\,+\,z^2}}\,+\,1\,-
2\nu\,\,\Bigr]
\label{extratwelve}
\eeq
where $z_{\infty}$ is a constant of integration. It can be verified that this
function satisfies the second-order differential equation (\ref{eqdiff1}),
as required to minimise the energy.

The behaviour of the solution (\ref{extratwelve}) has been analyzed in
detail in
\cite{ramallo}. It was found there that, for regions close to the
horizon, the embedding represented by the function $z(\rho)$ coincides with
the solutions found for $a=0$. However, for the asymptotic region
$\rho\rightarrow\infty$, the behaviour of both types of solutions is completely
different. Indeed, it follows from (\ref{extrathree}) and
(\ref{extrafour})  that for $\rho\rightarrow\infty$ and $z$ fixed (which
corresponds to $\theta\rightarrow \pi/2$), $dz /d\rho$ decreases as
$\rho^{-3}$, which implies that $z\rightarrow$constant as
$\rho\rightarrow\infty$, {\it i.e.} the asymptotic shape of the brane is just a
plane. This constant asymptotic value of $z$ is precisely the integration
constant $z_{\infty}$, as can be verified directly from (\ref{extratwelve}).
Notice that this is very natural from the physical point of view since our
metric is asymptotically flat and one expects the minimal energy configuration
of a tensile object in a flat metric to be such that the object is not bent at
all.

Under certain conditions our solution (\ref{extratwelve}) contains a tubular
region which has one of its ends near the horizon and can
locally be described by the equation $\rho=$constant. Since
$|dz/d\rho|$ is then large, these portions of the brane are not
located in the asymptotic region $\rho\rightarrow\infty$ in which, as
we argued above, $dz/d\rho\rightarrow 0$. These tubes therefore
spread out and approach the plane $z=z_{\infty}$ as one
moves away from the near-horizon region. Detailed study \cite{ramallo}
shows that for large
$|z_{\infty}|$ the transition between these two regimes is very fast. Notice
that these spikes, which have infinite length in the near-horizon
approximation, have actually a finite length (of the order of
$|z_{\infty}|$) in the full geometry. In this sense one can say
that the full metric regularizes the singularities of the near-horizon
description. Moreover,
it can be verified that the energy of the tubes per unit surface equals that
expected for a bundle of M2-branes. From this analysis
we get a picture of the HW effect in M-theory in which M5-branes are connected
by M2-branes which, from the worldvolume point of view, can be regarded as
wormholes connecting the test M5 brane to the horizon of the metric created by
the background M5-branes.

\subsection{Supersymmetry}

We will now show that configurations saturating the bound (\ref{Mb}) preserve
1/4 of the bulk supersymmetry. Our starting point is again (\ref{introa}).
The explicit form of $\Gamma_\kappa$ for the M5-brane can be found in
\cite{sorokinb}. Omitting some terms which are manifestly zero for the
configurations we consider, and passing to the temporal gauge for the PST
scalar, we have
\be
\sqrt{-\det(h+\7H)}\, \Gamma_\kappa = {1\over 2}\sqrt{-\det h}\,
\Gamma^{\underline{0}}\, \Gamma_{IJ}\7H^{IJ}- {1\over 5!}U^{-1/6}
\Gamma^{\underline{0}}\epsilon^{I_1\ldots I_5}\Gamma_{I_1\ldots I_5}\, .
\ee
Again $\Gamma_{IJ\cdots}$ are antisymmetrized products of the (reducible)
worldvolume gamma matrices $\Gamma_I=\6_I X^m E_m{}^{\underline{m}}
\Gamma_{\underline{m}}$, where $\Gamma_{\underline m}$ are the constant
D=11 Dirac matrices and $E_m{}^{\underline m}$ is the spacetime elfbein.
In the static gauge and for static configurations with $X^1=X^2=X^3=X^4=0$
we have
\bea
\Gamma_0 &=& U^{-\,1/3}\Gamma_{\underline 0} \nonumber \\
\Gamma_i &=& U^{1/3}\,r\gamma_i + U^{1/3}\6_i r \Gamma_{\underline{\natural}}
\, .
\eea
In this expression we have introduced the matrix
\be
\gamma_i = e_i{}^{\underline i}\Gamma_{\underline i}
\ee
where $e_i{}^{\underline i}$ is the $S^4$ vierbein. Thus
\be
\{\gamma_i,\gamma_j\}= 2\bar g_{ij}\, .
\ee
For the M5-brane background, the Killing spinors take the form
\be
\chi = U^{-{1\over 12}}\epsilon
\ee
where $\epsilon$ is covariantly constant on $\bE^{(1,5)}\times
\bE^5$ subject to a 1/2 supersymmetry breaking condition imposed by the
background, which we discuss below. The condition for preservation of
supersymmetry (\ref{introa}) is thus reduced to
\be\label{scon}
\Gamma_\kappa \epsilon = \epsilon\, .
\ee
In polar coordinates for $\bE^5$, with $\Gamma_{\underline{\hat 4}}
\equiv \Gamma_{\underline \t}$, we have
\be
\epsilon = \,e^{{\t\over
2}\,\Gamma_{\underline{\natural\t}}}\,\prod_{\8{\j}=1}^3
e^{-{\t^{\8{\j}} \over 2}\,
\Gamma_{\underline{\8{\j}{\8{\j}+1}}}}\,\,
\epsilon_0 \quad ,
\ee
where $\epsilon_0$ is a constant spinor satisfying
\be
\Gamma_{\underline {6789\natural}}\epsilon_0=\epsilon_0\, .
\ee
There are again additional Killing spinors in the near-horizon limit ($a=0$)
but these play no role in our analysis for the same reason as in the D5-brane
case.

The supersymmetry condition (\ref{scon}) gives
\bea\label{susycond}
& U^{-1/3}\sqrt{\det\left(g_{ij}+K_iK_j\right)} \epsilon =
\left[\Delta_4 r\Gamma_{\underline{0}} \gamma_*
-\Pi^i\6_i r \Gamma_{\underline{05\natural}}\right.
 \nonumber \\
& \left.-\gamma_i\Gamma_{\underline{\natural}}\left(
\Pi^i r \Gamma_{\underline{05\natural}} +
U\,r^3\,\sqrt{\det \5g}\, \5g^{ij}\6_j r
\Gamma_{\underline{0}} \gamma_* \right)\right]\epsilon
\eea
where
\be
K_i = {1\over \sqrt{\det g}}\, g_{ij}\Pi^j
\label{defi}
\ee
and
\be
\gamma_* = \Gamma_{\underline{56789}}\, .
\ee
To solve this equation we will impose the conditions
\be
\Gamma_{\underline{0}} \gamma_*\epsilon_0 = \epsilon_0 \, ,\qquad
\Gamma_{\underline{05\natural}}\epsilon_0 = \epsilon_0 \, .
\ee
which are the conditions expected for an $M5$-brane with an $M2$-brane ending
on it. One can then deduce the following identities :
\bea
\Gamma_{\underline{0}} \gamma_*\epsilon & = & (\cos\t -
\Gamma_{\underline{\natural\t}}\sin\t)\epsilon \\
\Gamma_{\underline{05\natural}}\epsilon & = & (\cos\t -
\Gamma_{\underline{\natural\t}}\sin\t)\epsilon \, .
\eea
The matrix multiplying $\epsilon$ on the right hand side of
(\ref{susycond}) can be written as
\bea
& \Delta_4\left(r\sin\t\right)' -\Pi^i\6_i \left(r\,\cos\t\right)
+ \Gamma_{\underline{\natural\t}}\left[\Delta_4\left(
r\cos\t\right)' +\Pi^i\6_i \left(r\sin\t\right)\right] \nonumber \\
& + \Gamma_{\underline{\natural}}\gamma_{\8{\i}}\left(\Delta_4
\5g^{\8{\i}\8{\j}}\6_{\8{\j}}r\cos\t +\Pi^{\8{\i}}r\cos\t
\right) +\gamma_{\8{\i}}\Gamma_{\underline{\t}}\left(\Delta_4
\5g^{\8{\i}\8{\j}}\6_{\8{\j}}r \sin\t +\Pi^{\8{\i}}r\sin\t
\right)
\label{e10}
\eea
Requiring the coefficient of $\Gamma_{\underline{\natural}}
\gamma_{\8{\i}}$ to vanish, we find that
\be
\Pi^{\8{\i}}=-\Delta_4{\5g^{\8{\i}\8{\j}}\6_{\8{\j}}r \over r}
\label{BPS1}
\ee
Requiring the coefficient of $\Gamma_{\underline{\natural\t}}$ to vanish
and using (\ref{BPS1}) we then find that
\be
\Pi^\t (r\sin\t)' = -\Delta_4\left[
(r\cos\t)' - {\5g^{\8{\i}\8{\j}}\6_{\8{\i}}r\6_{\8{\j}}r 
\sin\t \over r}\right]\, .
\label{BPS2}
\ee
Notice that (\ref{BPS1}) and (\ref{BPS2}) are entirely equivalent
to the equations saturating the energy bound (\ref{bpse1}) and
(\ref{bpse2}). It is straightforward to check that when (\ref{BPS1}) is
satisfied, the coefficient of $\gamma_{\8{\i}}\Gamma_{\underline{\t}}$
vanishes identically, so (\ref{e10}) is satisfied as a consequence
of (\ref{BPS1}) and (\ref{BPS2}) provided that
\be
U^{-1/3}\,\sqrt{\det\left(g_{ij}+K_i K_j \right)} =
{\Delta_4\,r^2 \over \left(r\sin\t\right)'}\left(
1+ {\5g^{ij}\6_i r\6_j r \over r^2}\right)\, .
\label{result}
\ee
Since
\be
\det\left(g_{ij}+K_i K_j \right)=(\det g)\left(1+g^{ij}K_i K_j\right)\,
\ee
one can use (\ref{defi}) to rewrite the left hand side of  (\ref{result})
as
\be
U^{-1/3}\,\det g + U^{1/3}\left[r^2\Pi^i\Pi^j\5g_{ij} +
\left(\Pi^i\6_i r\right)^2\right]\,
\ee
which can be shown to satisfy (\ref{result}) when (\ref{BPS1}) and
(\ref{BPS2}) are satisfied. We have thus shown that equations (\ref{BPS1})
and (\ref{BPS2}) are sufficient for preservation of $1/4$ supersymmetry.
As we saw earlier, when these equations are combined with the Bianchi identity
they become equivalent to the second order equation (\ref{eqdiff1}). Any
solution of this equation is therefore 1/4 supersymmetric.

\subsection{Multi-angle M5 solutions with 1/4 supersymmetry}

We shall proceed as in subsection 2.3 to find more general solutions
to equation (\ref{eqdiff1}) in the near horizon region. Our starting point will
be the $SO(4)$ invariant  solution for $\nu\neq 0$, which is 
~\cite{callan2, ramallo} :
\begin{equation}
r_\nu(\t) =C{\left(\Lambda(\t)\right)^{1/2}\over \sin\t}
\end{equation}
where $\Lambda(\t)=2(1-2\nu-\cos\t)$. This function satisfies the 
following identity:
\begin{equation}\label{miracle}
{d\over d\t}\left({r_\nu\sin^2\t\over (r_\nu\sin\t)^\prime}\right)=
2\sin\t \, ,
\end{equation}
Using the separation of variables ansatz 
\begin{equation}
r(\theta,\theta_{\8{\i}})=r_{\nu}(\theta)\prod_{\8{\i}}\delta_{\8{\i}}
(\theta_{\8{\i}})\, ,
\label{ans1}
\end{equation}
we find that (\ref{eqdiff1}) reduces to
\begin{eqnarray}\label{g1}
0 & = & \sin\t\sin\t_2\left[2\sin\t_3\cos\t_3\psi_3 + \sin^2\t_3\psi_3^\prime
\right] \nonumber \\
& & +\sin\t\left[\cos\t_2\psi_2 + \sin\t_2 \psi_2^\prime\right]
+ {\sin\t\over \sin\t_2}\psi_1^\prime \nonumber \\
& & -\sin^2\t_3\sin\t_2 {d\over d\t}\left({r_\nu\sin^2\t\over
(r_\nu\sin\t)^\prime}\right)\left[\psi_3^2 + {\psi_2^2\over \sin^2\t_3}+
{\psi_1^2\over \sin^2\t_3\sin^2\t_2}\right] \, ,
\end{eqnarray}
where $\psi_{\8{\i}}(\t_{\8{\i}})$ are defined as in subsection 2.3.
Using the identity (\ref{miracle}), this can be shown to be
equivalent to
\begin{eqnarray}\label{g2}
0 & = & \sin\t_2\left[2\sin\t_3\cos\t_3\psi_3 + \sin^2\t_3\psi_3^\prime
- 2\sin^2\t_3\psi_3^2\right] \nonumber \\
& & +\left[\cos\t_2\psi_2 + \sin\t_2\psi_2^\prime - 2\sin\t_2\psi_2^2\right]
\nonumber \\
& & +{1\over \sin\t_2}\left[\psi_1^\prime -2\psi_1^2\right] \, .
\end{eqnarray}
We can solve this equation as before by choosing the $\psi$-functions to
satisfy the Bernouilli equations
\footnote{Analogous generalizations to the ones pointed out in section
2.3 also apply in this case.} 
\begin{eqnarray}
\psi_1' +2\cot\t_1\psi_1 & = & 2\psi^2_1
\label{m1} \\
\psi_2' + \cot\t_2\psi_2 & = & 2\psi^2_2
\label{m2} \\
\psi_3' & = & 2\psi_3^2\, .
\label{m3}
\end{eqnarray}
In addition to the trivial solution $\psi_{\hat{\i}}(\theta_{\hat{\i}})=0$, 
these have the non-trivial solutions
\begin{eqnarray}
{1\over \psi_1} & = & c_1\sin^2\t_1 + 2\sin\t_1\cos\t_1
\label{n1} \\
{1\over \psi_2} & = &
c_2\sin\t_2 \label{n2}-2\sin\t_2 \log\tan {\t_2\over 2} \\
{1\over \psi_3} & = & c_3 -2\t_3 \, .
\label{n3}
\end{eqnarray}
The singularity structure of the new M5-brane solutions is similar to what we
found earlier for the $D5$-brane. Any non-constant function $\delta_{\hat{\i}}$
will have a branch cut singularities at some angle $\alpha_{\hat{\i}}$ of the 
type
\begin{equation}
\delta_{\8{\i}} \sim {1\over (\t_{\8{\i}} - \alpha_{\8{\i}})^{1\over 2}}\, .
\end{equation}
To compute the energy of the new solutions we would need the conjugate momenta
to the angles. These are
\begin{eqnarray}
\Pi^3 & = & -R^3\sin\t\,\sqrt{det\,^3g} \,\psi_3(\t_3)
\label{cmm1}\\
\Pi^2 & = & -R^3\sin\t\,\sqrt{det\,^3g} \, {\psi_2(\t_2)\over \sin^2\t_3}
\label{cmm2} \\
\Pi^1 & = & -R^3\sin\t\,\sqrt{det\,^3g} \,{\psi_1(\t_1)\over
\sin^2\t_3\sin^2\t_2} \label{cmm3} \\
\Pi^\t & = & \Pi(\t)\,\sqrt{det\,^3g} +4R^3\Lambda(\t)\,
\,\sqrt{det\,^3g}\left[
(\psi_3(\t_3))^2 + {(\psi_2(\t_2))^2\over \sin^2\t_3} \right. \nonumber \\
& & \left. + {(\psi(\t_1))^2\over \sin^2\t_3\sin^2\t_2}\right] \, .
\end{eqnarray}
The corresponding topological charge is
\begin{eqnarray}
{\cal Z}_4 & = &  \left[{\cal Z}(\t)\,\sqrt{det\,^3g} + 4R^3
{r_0(\t)\over \sin\t}\,\,\Lambda(\theta)\,
\sqrt{det\,^3g}\right.
\nonumber \\
& & \left. \left((\psi_3(\t_3))^2+ {(\psi_2(\t_2))^2\over \sin^2\t_3} +
{(\psi(\t_1))^2\over \sin^2\t_3\sin^2\t_2}\right)\right]
\prod_{\hat {\i}}\delta_{\hat {\i}}(\t_{\hat {\i}}) \, .
\label{tc2}
\end{eqnarray}
It can be verified that the $\Pi$'s satisfy  (\ref{bity1}).
It is again remarkable that the new contributions to the topological
charge $\psi_{\8{\i}}$ are positive whenever the $SO(4)$
invariant configuration is positive (for the full range of $\theta$ when
$\nu=0$). As for the D5-D3 configuration, ${\cal Z}_4$ can be rewritten
as a divergence, ${\cal Z}_4 = \partial_i {\cal Z}_4^i$, at least in the near 
horizon limit. Taking this limit in (\ref{div2}) we find that 
\bea
{\cal Z}_4^\t & = & -\Pi^\t\,r\cos\t +R^3\,\sin^4\t\,\sqrt{det\,^3g}\,r
\nonumber \\
{\cal Z}_4^{\8{\i}} & = & -\,\Pi^{\8{\i}}\,r\cos\t\, .
\label{div22}
\eea

\section{Discussion}

A universal feature of supersymmetric p-branes is the appearance in the action
governing their dynamics of a Wess-Zumino (WZ) term. In the simplest cases this
term just describes the coupling of the brane to a (p+1)-form potential of the
supergravity background, and the physics associated with this coupling has been
extensively investigated in the past. For both D-branes and the M5-brane this
coupling is only the first term in an expansion of the WZ term in powers of the
field strength of a worldvolume gauge field. In certain backgrounds the first
non-leading term in this expansion provides a source term for the
worldvolume gauge field which, as a consequence, cannot vanish. The baryonic
brane configurations discussed here, and in previous work, provide examples of
the physics associated with this phenomenon. In previous work a 1/4
supersymmetric baryonic D5-brane with $SO(5)$ symmetry was found and shown to
saturate an energy bound, and similar results have been established for other 
D-branes. One result of this paper is an extension of these results to the
M5-brane. In this case the baryonic brane provides a worldvolume realization of
the baryon string-vertex of the $(2,0)$ superconformal field theory which,
according to the adS/CFT correspondence, lives on the boundary of the
$adS_4\times S^7$ near-horizon background of the supergravity M5-brane.

Another purpose of this paper has been to find the {\sl general} conditions
implied either by saturation of a Bogomol'nyi-type energy bound or by 1/4
supersymmetry, for both baryonic D5-branes and baryonic M5-branes. We have 
found
these conditions, and we have found that they coincide, i.e. configurations
saturating the bound are 1/4 supersymmetric and vice-versa. At least, this is
formally true in that both conditions yield the same second-order partial
differential equation for a function $r$ on $S^5$ (in the D5-brane case) or
$S^4$ (in the M5-brane case). Of course, the equations for the two cases, D5
and M5 are different, but they are broadly similar. Particular solutions are
easily found in the D5-brane case via an $SO(5)$-invariant ansatz. The
condition for 1/4 supersymmetry then reduces to one found previously, and the
preciously found baryonic brane solution is recovered. We have shown how a
similar $SO(4)$-invariant ansatz yields an explicit baryonic M5-brane 
solution. 

In going beyond these simple solutions of high symmetry one runs into the
difficulty that a given local solution of the 1/4 supersymmetry condition
will not generally be well defined on the 5-sphere (4-sphere in the M5
case, but we concentrate on the D5 case in the following discussion). Of 
course,
no solution can really be well-defined on the 5-sphere because the D5-brane
carries a non-zero charge but, for reasons that we have explained in the
introduction, one might suppose that the singularities could be confined to
points, which could then be interpreted as the endpoints of N
strings. The $SO(5)$-invariant solution has a single point singularity which
has been interpreted as the coincident endpoints of $N$ parallel strings. 
However, although we have found many new 1/4 supersymmetric solutions, none of
them has singularities of this type; in fact they have branch cuts. We suspect
that this is due to the restricted nature of the ansatz that we have used
to find these solutions. The full equation for 1/4 supersymmetry is not at all
simple and one cannot expect to find all its solutions explicitly. Thus, our
inability to find solutions with specified singularities is in no way an
indication that they do not exist, although it is also true that we cannot
prove that they do. The point is an important one because it has implications 
for potential limitations of the worldvolume approach to spacetime physics, but
its full resolution must await future studies.

\medskip
\section*{Acknowledgments}
\noindent
PKT thanks the ECM of the University of Barcelona for hospitality.
JS thanks the ITP of the University of Hannover for hospitality. JS is
supported by a fellowship from Comissionat per a Universitats i Recerca
de la Generalitat de Catalunya. This work was supported in part by
AEN98-0431 (CICYT), GC 1998SGR (CIRIT). JG and JS would like to thank
J. Sol\`a-Morales for useful discussions.
AVR is grateful to J. M. Sanchez de Santos for discussions and early
collaboration on some of the topics studied in this paper. The work of AVR
is supported in part by CICYT under grant AEN96-1673 and  by DGICYT
under grant PB96-0960 and by the European Union grant ERBFM-RXCT960012.


\end{document}